\documentclass[12pt]{article}
\usepackage{hyperref}
\usepackage{amsmath}
\usepackage{graphicx}
\usepackage{amssymb} 
 \usepackage{xcolor}
\definecolor{myblue}{rgb}{0.14,0.11,0.49}
\definecolor{myred}{rgb}{0.74,0.22,0.15}
\definecolor{mygreen}{rgb}{0.05,0.52,0.42}
\definecolor{myyellow}{rgb}{0.96,0.92,0.13}
\definecolor{myorange}{rgb}{1,0.61,0.36}
\definecolor{mypurple}{rgb}{0.71,0.02,1}
\definecolor{noir}{gray}{0.} 
\newcommand{\Couleur}[1]{\textcolor{noir}{#1}}
\newcommand{\Mat}[1]{{{\boldsymbol{#1}}}}
\newcommand{\abs}[1]{\left\vert#1\right\vert}
\def\be{\begin{equation}}
\def\ee{\end{equation}}
\def\bea{\begin{eqnarray}}
\def\eea{\end{eqnarray}}
\def\bi{\begin{itemize}}
\def\ei{\end{itemize}}
\def\noi{\noindent}
\def\dd{\mathrm{d}}

\date{}

\title{Should there be a spin-rotation coupling for a Dirac particle?}

\author{Mayeul Arminjon\\
\small\it Laboratory ``Soils, Solids, Structures, Risks'', 3SR\\ \small\it (CNRS and Universit\'es de Grenoble: UJF, Grenoble-INP),\\\small\it BP 53, F-38041 Grenoble cedex 9, France.}

\begin{document}
\maketitle

\begin{abstract} 
\noi It was argued by Mashhoon that a spin-rotation coupling term should add to the Hamiltonian operator in a rotating frame, as compared with the one in an inertial frame. For a Dirac particle, the Hamiltonian and energy operators H and E were recently proved to depend on the tetrad field. We argue that this non-uniqueness of H and E really is a physical problem. We compute the energy operator in the inertial and the rotating frame, using three tetrad fields: one for each of two frameworks proposed to select the tetrad field so as to solve this non-uniqueness problem, and one proposed by Ryder. We find that Mashhoon's term is there if the tetrad rotates as does the reference frame --- but then it is also there in the energy operator for the inertial frame. In fact, the Dirac Hamiltonian operators in two reference frames in relative rotation, but corresponding to the same tetrad field, differ only by the angular momentum term. If the Mashhoon effect is to exist for a Dirac particle, the tetrad field must be selected in a specific way for each reference frame.
\end{abstract} 

\section{Introduction}\label{Intro}

In a reference frame that has a uniform rotation with respect to an inertial frame, the angular momentum ${\bf L}$ of a particle is coupled with the rotation of the frame, in the sense that the Hamiltonian function or operator of the particle differs from its expression in the inertial frame by the term $-\Mat{\omega.} {\bf L}$. (Here, $\Mat{\omega}$ is the constant rotation velocity vector.) In the non-relativistic framework (also in the presence of Newtonian gravitation), this is exact for the classical Hamiltonian function as well as for the quantum Hamiltonian operator --- when, to define the latter, one considers a scalar particle without spin \cite{WernerStaudenmannColella1979,A41}. (In a relativistic framework, for a particle without spin obeying the Klein-Gordon equation, the Hamiltonian operator in a rotating frame may have other terms involving ${\bf L}$, depending on the model metric which is considered \cite{Kuroiwa-et-al1993, MorozovaAhmedov2009}.) Therefore, if one considers that the spin of a quantum particle is expressing some kind of internal rotation, he may conjecture that also the spin might couple with the rotation of the reference frame. This could even be regarded \cite{Mashhoon1995} as a natural consequence of the fact that the total angular momentum operator is the sum of the orbital momentum and the spin. Thus, it was argued by Mashhoon \cite{Mashhoon1988} that a ``spin-rotation coupling" term of the form 
\be\label{Mashhoon term}
\delta \mathrm{H_{SR}}=-\gamma_\mathrm{L} \,\Mat{\omega .}{\bf S}
\ee
should add, in a uniformly rotating frame, to a quantum Hamiltonian H of relativistic quantum mechanics. Here, $\gamma_\mathrm{L} $ is the Lorentz factor corresponding to the velocity, with respect to the inertial frame, of the local observer attached with the rotating frame, and ${\mathbf S} \equiv \frac{1}{2}\hbar\Mat{\sigma}$, where $\Mat{\sigma}$ denotes the space ``vector" made with the Pauli matrices $\sigma ^j \ (j=1,2,3)$. The form of H was left free by Mashhoon, who got the additional term (\ref{Mashhoon term}) from an assumption about the transformation of H and the wave function from the inertial frame to the rotating one. Later on, a similar term: 
\be\label{HehlNi term}
\delta \mathrm{H_{SR}}'=-\,\Mat{\omega }'{\bf .S}
\ee 
with $\Mat{\omega }'$ the ``proper rotation", was predicted by Hehl \& Ni \cite{HehlNi1990} to occur in the Hamiltonian of a particle obeying specifically the standard form \cite{BrillWheeler1957+Corr,ChapmanLeiter1976} of the (generally-)covariant Dirac equation (``Dirac-Fock-Weyl" equation or DFW for short). To write the latter explicitly, one needs to define a coordinate system and an (orthonormal) tetrad field. That prediction was obtained for a general situation in which an observer moves with a proper acceleration and a proper rotation, yet a particular tetrad field was chosen, ``which behaves as a rotating Fermi-Walker-transported reference frame" \cite{HehlNi1990}. Still a similar prediction was got, also from the DFW equation but in the case of uniform rotation, by Cai \& Papini \cite{CaiPapini1991} who used another ``rotating tetrad". In view of these results, and since the Dirac equation is the relevant one to describe spin half particles, the spin-rotation coupling is usually considered as a theoretically well established fact. It seems that it is too small to be experimentally tested yet \cite{A41}, but it has been argued that it may have been indirectly detected \cite{Mashhoon1995}, although the argument is not very straightforward.\\

Until recently, the choice of the tetrad field has been assumed to be entirely neutral, because the Lagrangian of the standard covariant Dirac equation is invariant under a change of the tetrad field, hence the DFW equations obtained with any two different tetrad fields are equivalent \cite{BrillWheeler1957+Corr,ChapmanLeiter1976}. (This is true in a topologically-simple spacetime \cite{Isham1978}.) However, it has been observed by Ryder \cite{Ryder2008} that, in the archetypical case of uniform rotation in the Minkowski spacetime, the spin-rotation coupling term may be present or absent, depending on the choice of the tetrad field. Even more recently, it has been proved that, in a general reference frame in a general spacetime, the Hamiltonian operator H associated with the covariant Dirac equation is not unique \cite{A43}. (This is true for the DFW equation as well as for all alternative forms of the covariant Dirac equation considered in Refs. \cite{A43, A45}, for which the gauge freedom is even larger than for the DFW equation.) In loose terms, the reason for this non-uniqueness is as follows: H is got by rewriting the wave equation in a form adapted to a particular reference frame; now, for the covariant Dirac equation, the Dirac $\gamma ^\mu $ matrices and their admissible changes are allowed to depend on the spacetime position; it follows that rewriting the covariant Dirac equation in a form adapted to a particular reference frame does not generally commute with changing the $\gamma ^\mu $ matrices in that equation. It has also been proved \cite{A43} that the {\it energy operator} E is not unique, either. That operator E is equal to the Hermitian part of the Hamiltonian operator H for the relevant scalar product (hence it coincides with H when H is Hermitian) and has the other important property that its mean value is the field energy \cite{Leclerc2006,A48}. Thus it is the energy operator E that is relevant to the Mashhoon effect. 
The spectrum of E, that is the Dirac energy spectrum, is not unique either. Instead, each of H, E, and the spectrum of E, depend on the choice of the tetrad field, or more generally of the field of Dirac matrices \cite{A43}. Thus, contrary to a widely spread belief, the gauge invariance of the Lagrangian of the DFW equation --- i.e., its invariance under any smooth change of the tetrad field --- does not guarantee that all physically-relevant objects are also gauge-invariant.\\
 
In contradiction with the criticism \cite{GorbatenkoNeznamov2013}, that non-uniqueness does not regard merely the form of H and E. Indeed, what has been proved \cite{A43} is the physical inequivalence of the Hamiltonians (and the energy operators) corresponding with different choices of the tetrad field. See Ref. \cite{A50} for another detailed proof of that point using precisely the concepts of a unitary transformation and of the mean value of an operator, invoked in Ref. \cite{GorbatenkoNeznamov2013}. Let us now emphasize that this physical non-uniqueness of H and E really is a problem. \\

{\bf i}) The work \cite{A48}, App. A, provides a detailed justification for using first-quantized covariant Dirac theory, that is, {\it quantum mechanics of the covariant Dirac equation}, instead of quantum field theory (QFT), in the context of the existing experiments on quantum particles in the gravitational field. In short: curved-spacetime QFT applied to the Dirac field, in its current state, does not allow one to make unambiguous predictions about the COW effect, the Sagnac effect, the quantization of the energy levels in the gravitational field, all of those three effects having been confirmed by experiments, and which are the only available experiments relative to the gravity-quantum coupling. Nor does the current state of QFT allow one to make predictions regarding the Mashhoon effect --- which is foreseen to become measurable, has been widely discussed in the literature, and is the precise subject of this paper.\\

{\bf ii}) There also, the issue of the classical {\it energy-momentum tensor} is discussed in relation with the non-uniqueness problem. It is shown that the {\it canonical} energy-momentum tensor $t^\mu _{\ \, \nu }$ is the one for which the ``field energy", i.e. the space integral of $t^0 _{\ \, 0 }$, is equal to the mean value $\langle \mathrm{E} \rangle $ of the energy operator $\mathrm{E}$. Thus, the canonical tensor $t^\mu _{\ \, \nu }$ is the one that is relevant to quantum mechanics and to these experiments --- but $t^\mu _{\ \, \nu }$ is not gauge-invariant: this can be checked directly on its expression and results also from the foregoing equality, since $\mathrm{E}$ and $\langle \mathrm{E} \rangle $ are not gauge-invariant. On the other hand, Hilbert's energy-momentum tensor, say $T^\mu _{\ \, \nu }$, is gauge-invariant, but the space integral of $T^0 _{\ \, 0 }$ is hence not equal to the mean value $\langle \mathrm{E} \rangle $ of the energy operator $\mathrm{E}$. 
Hence, $T^\mu _{\ \, \nu }$ is not relevant to quantum mechanics. Anyway, in the rather vast literature on quantum mechanics of the DFW equation (see e.g. Refs. \cite{HehlNi1990}--\cite{ChapmanLeiter1976}, \cite{Leclerc2006}--\cite{GorbatenkoNeznamov2013}, \cite{Parker1980}--\cite{HuangParker2009}), the energy-momentum tensor (be it $t^\mu _{\ \, \nu }$ or $T^\mu _{\ \, \nu }$) is rarely even mentioned, except for Refs. \cite{BrillWheeler1957+Corr,Leclerc2006}. In any case, to our knowledge, that tensor has never been used in a calculation that have a definite relationship to the outcomes of the existing or foreseen experiments testing the effects of the gravity-quantum coupling, mentioned at point ({\bf i}) above.

{\bf iii}) The non-uniqueness problem is there already in the case of an inertial frame in the Minkowski spacetime \cite{A47}, and this is also true in the presence of an external electromagnetic field.
\footnote{
\label{em-case}\
In that case, the r.h.s. of the ``free" Dirac equation (\ref{Dirac-normal}) is augmented with the term $-iq\gamma ^\mu V_\mu \Psi $, with $q$ the electric charge and $V_\mu $ the four-potential. Thus the ``free" Hamiltonian H is replaced by $\mathrm{H}_{\mathrm{em}}=\mathrm{H}+q(V_0\,{\bf 1}_4+V_j\,\alpha ^j)$, where $\alpha ^j\equiv \gamma ^0\gamma ^j/g^{00}$, as is the case \cite{A40} for Dirac's original equation. It follows that, after a local similarity transformation $S$, after which H becomes $\widetilde{\mathrm{H}}$, the complete Hamiltonian H$_{\mathrm{em}}$ becomes $\widetilde{\mathrm{H}_{\mathrm{em}}}$, with  $\widetilde{\mathrm{H}_{\mathrm{em}}}-S^{-1}\mathrm{H}_{\mathrm{em}}S=\widetilde{\mathrm{H}}-S^{-1}\mathrm{H}S$. We get similarly for the energy operator: 
$\widetilde{\mathrm{E}_{\mathrm{em}}}-S^{-1}\mathrm{E}_{\mathrm{em}}S=\widetilde{\mathrm{E}}-S^{-1}\mathrm{E}S$, whence for any state $\Psi $ and the corresponding state after application of $S$, $\widetilde{\Psi }\equiv S^{-1}\Psi $ [noting $(\,\mid \,)$ and $(\,\,\widetilde{\mid} \,  \,)$ the scalar products before and after application of $S$]:
\be\label{Etilde-Ebreve-em}
(\widetilde{\Psi }\,\widetilde{\mid} \,  \widetilde{\mathrm{E}_{\mathrm{em}}}\widetilde{\Psi })-(\Psi \mid \mathrm{E}_{\mathrm{em}}\Psi )=(\widetilde{\Psi }\,\widetilde{\mid} \,  \widetilde{\mathrm{E}}\widetilde{\Psi })-(\Psi \mid \mathrm{E}\Psi ),\ \mathrm{or}\ \langle \widetilde{\mathrm{E}_{\mathrm{em}}}\rangle -\langle \mathrm{E}_ {\mathrm{em}} \rangle = \langle \widetilde{\mathrm{E}}\rangle-\langle \mathrm{E} \rangle.
\ee
\{We use the fact that $(\widetilde{\Psi }\,\widetilde{\mid} \,  S^{-1}\mathrm{E}S \widetilde{\Psi })=(\Psi \mid \mathrm{E}\Psi )$ \cite{A50}.\} Hence, the non-uniqueness of the operators H$_{\mathrm{em}}$ and E$_{\mathrm{em}}$ and that of the spectrum of E$_{\mathrm{em}}$ appear in strictly the same way as in the case of the ``free" Dirac equation, whether the spacetime is curved or not.
}
The classical discussion of the hydrogen-type atoms, which is based on the quantum-mechanical Hamiltonian/energy operator and its spectrum, therefore cannot be done if one uses the DFW equation with its gauge freedom, instead of using Dirac's original equation valid only in Cartesian coordinates \cite{A48}. I.e.: {\it the current theory based on the DFW equation cannot determine the energy levels of the hydrogen atom.} See Eq. (\ref{bar A-explicit}) below. This illustrates in a dramatic way the physical relevance of the non-uniqueness problem. \\

{\bf iv}) Finally, the  principle according to which ``physical observables are gauge invariant" cannot discard the energy operator, because this is the most important quantum-mechanical observable --- as is confirmed by point ({\bf iii}) above. What this principle tells us in that instance is that {\it we have to restrict the gauge freedom:} here the freedom in the choice of the tetrad field. \\

\vspace{2mm}
\noi That non-uniqueness problem makes it plausible that a spin-rotation coupling term could be unambiguously defined only if the choice of the tetrad field were restricted in some consistent way. Note that the derivations which lead to the presence of a spin-rotation coupling term for a Dirac particle are based on choosing a tetrad that is itself rotating more or less like the reference frame \cite{HehlNi1990,CaiPapini1991}, as is also the case for Ryder's first tetrad \cite{Ryder2008}. Whereas, Ryder's second tetrad, which does not lead to the presence of this term, is indeed non-rotating in the sense of the Fermi-Walker transport \cite{Ryder2008}.\\

Two different frameworks have been proposed \cite{A48,A47} to restrict the choice of the tetrad field in such a way that the non-uniqueness problem \cite{A43} is proved to be solved:\\

 {\bf I.} With any orthonormal tetrad field $(u_\alpha )_{\alpha =0,...,3}$ that is ``adapted" to a given reference frame (in a sense to be precised in Sect. \ref{Prescriptions}), one may associate a unique rotation rate field $\Mat{\Xi }$, which is a spatial tensor field. A first way to solve the non-uniqueness problem is to fix that spatial tensor field $\Mat{\Xi }$ \cite{A47}. Two natural choices for this fixing are: {\it a}) $\Mat{\Xi }=\Mat{\Omega }$, where $\Mat{\Omega }$ is the rotation-rate field of the reference frame itself \cite{Cattaneo1958,Weyssenhoff1937,A47}; and {\it b}) $\Mat{\Xi }=\Mat{0}$. These two choices lead to non-equivalent Hamiltonians, thus represent two different solutions to the non-uniqueness problem.\\

{\bf II.} A third solution is available \cite{A48} when the spacetime metric $\Mat{g}$ can be put in the following diagonal space-isotropic form:
\be\label{isotropic-diagonal}
(g_{\mu \nu })=\mathrm{diag}(f,-h,-h,-h), \qquad f>0,\ h>0
\ee
in a suitable coordinate system $(x^\mu )$. That other solution consists in choosing the ``diagonal tetrad" in that coordinate system, i.e.,
\be\label{diagonal-tetrad}
u_\alpha  \equiv \delta _\alpha ^\mu \,\partial _\mu /\sqrt{\abs{d_\mu }}, \qquad d_0\equiv f, \quad d_1=d_2=d_3 \equiv -h.
\ee
So the two frameworks lead to three different prescriptions for uniqueness.\\

The aim of this work was to compare these two frameworks for the cases of both an inertial frame and a uniformly rotating frame in the Minkowski spacetime, with special attention to the presence or absence of the Mashhoon term. Section \ref{Operators} will recall the definition and the general form of the Dirac Hamiltonian operator $\mathrm{H}$ in a general spacetime, and Section \ref{Dependence} will distinguish between the dependences of $\mathrm{H}$ on the reference frame and on the tetrad field. Section \ref{Prescriptions} will give some additional details about the three different prescriptions outlined above. Section \ref{inertial/rotating}, which contains the main new results of this paper, will apply the foregoing to the target situation. In the Minkowski spacetime, the second framework leads one to select the ``Cartesian tetrad" and is very easy to put into practice. As this paper shall confirm, the first framework is much less easy to implement. So, instead of calculating exactly the predictions of each among the two variants of the first framework, we shall determine a tetrad field which closely approaches Variant {\it a}). 
We shall also test a rotating tetrad field proposed by Ryder \cite{Ryder2008}. In each case, i.e., in the two frames and for these three tetrad fields, we shall give the explicit expression of the energy operator. We shall finally find the general expression for the difference between the Hamiltonians corresponding to a given tetrad field, in two frames having a relative rotation.
 
\section{Dirac Hamiltonian operator in a general spacetime}\label{Operators}

The standard form of the covariant Dirac equation (the DFW equation) is written in a given coordinate system $(x^\mu )$ defined on the spacetime V (or on an open domain U therein):
\be\label{Dirac-normal}
\gamma ^\mu D_\mu\Psi=-iM\Psi \qquad (M\equiv mc/\hbar).
\ee
In this equation, $\gamma ^\mu $ is the field of the Dirac matrices; $\Psi $ is the column matrix made with the components $\Psi ^a \ (a=0,...,3)$ of the wave function $\psi $; and $D_\mu=\partial _\mu +\Gamma _\mu $ is the covariant derivative, where $\Gamma _\mu \ (\mu =0,...,3)$ are the connection matrices, which are $4\times 4$ complex matrices, just like the Dirac matrices $\gamma ^\mu $. On the other hand, $m$ is the rest-mass of the Dirac particles considered. The Dirac Hamiltonian operator is got by rewriting (\ref{Dirac-normal}) in the form of the Schr\"odinger equation and is explicitly \cite{A42}:
\be \label{Hamilton-Dirac-normal}
\mathrm{H} =  mc^2\alpha  ^0 -i\hbar c\alpha ^j D _j -i\hbar c\Gamma _0,
\ee
where
\be \label{alpha}
\alpha ^0 \equiv \gamma ^0/g^{00}, \qquad \alpha ^j \equiv \gamma ^0\gamma ^j/g^{00} \quad (j=1,2,3).
\ee
In contrast with the wave equation (\ref{Dirac-normal}), the Hamiltonian operator (\ref{Hamilton-Dirac-normal}) changes in a non-covariant way on a general change of the coordinate system. 
\footnote{\label{Covariance Psi}\
Nevertheless, the covariant Dirac equation being in particular covariant on a coordinate change, the evolutions of $\Psi $ calculated from $i\,\partial _t \Psi =\mathrm{H}\Psi $ in one coordinate system or in another one are equivalent. Specifically, for the DFW equation, $\Psi $ behaves as a scalar on any coordinate change \cite{BrillWheeler1957+Corr, ChapmanLeiter1976,A45}, thus we have simply $\Psi'((x'^\nu ))=\Psi((x^\mu ))$ --- with the restriction mentioned after Eq. (\ref{Minkowski in rotating Cartesian}) below.
}
This is true for any wave equation. However, the Hamiltonian operator transforms covariantly on a purely spatial change of the coordinates:
\be\label{purely-spatial-change}
x'^0=x^0,\quad x'^j=f^j((x^k)) \qquad (j,k=1,2,3).
\ee
In particular, for the DFW equation, $\Psi$ behaves as a scalar on  any coordinate change (Note \ref{Covariance Psi}). It follows easily \cite{A42} that the Dirac Hamiltonian is {\it invariant} after a change (\ref{purely-spatial-change}). \\

In the covariant Dirac equation (\ref{Dirac-normal}), as well as in the Hamiltonian operator (\ref{Hamilton-Dirac-normal}), the $\gamma ^\mu $ field is determined from the data of an orthonormal tetrad field $(u_\alpha )$. Decomposing the vectors $u_\alpha $ in the natural basis $(\partial _\mu )$: $u_\alpha =a^\mu_{\ \,\alpha}\,\partial_\mu$, one defines 
\be \label{flat-deformed}
 \gamma ^\mu = a^\mu_{\ \,\alpha}  \ \gamma ^{ \sharp \alpha},
\ee 
where $(\gamma ^{ \sharp \alpha})$ is any constant ``flat" set of Dirac matrices, i.e., one that is valid for the Minkowski spacetime in Cartesian coordinates \cite{BrillWheeler1957+Corr,ChapmanLeiter1976}. The definition (\ref{flat-deformed}) implies that the $\gamma ^\mu $ field transforms as a vector on a coordinate change (alone):
\be\label{gamma^mu vector}
\gamma'^\mu =L^\mu  _{\ \,\nu  }\gamma ^\nu ,\qquad L^\mu  _{\ \,\nu  }\equiv \frac{\partial x'^\mu }{\partial x^\nu } ,
\ee
which is well known. 
\footnote{\
There are alternative versions of the covariant Dirac equation in which the wave function is a complex vector field, for which case one may optionally decompose the wave function on the coordinate basis (the natural basis of the coordinate system) \cite{A45}. Taking this option means that the frame field on the spinor bundle coincides with the coordinate basis. Then $\Psi $ transforms as a vector and $(\gamma ^\mu) $ as a $(2\ 1)$ tensor \cite{A45}.
}
On the other hand, the connection matrices $\Gamma _\mu $ transform as a covector when one changes (only) the coordinate system:
\be\label{Gamma_mu covector}
\Gamma '_\mu = M^\nu _{\ \,\mu } \Gamma _\nu,\qquad M^\nu _{\ \,\mu } \equiv \frac{\partial x^\nu }{\partial x'^\mu } .
\ee
Using (\ref{gamma^mu vector}) and (\ref{Gamma_mu covector}), the invariance of the Hamiltonian operator H under a purely spatial change (\ref{purely-spatial-change}) is also easy to check directly on the explicit form (\ref{Hamilton-Dirac-normal}). We note that Eq. (\ref{Gamma_mu covector})$_1$ relates the matrices $\Gamma _\nu $ and $\Gamma '_\mu$ of any connection (on some vector bundle ${\sf E}$ with base V) when two different frame fields are chosen for the tangent bundle TV, say $(u_\nu )$ and $(u'_\mu)$ with $u'_\mu =M^\nu _{\ \,\mu }u_\nu $, even if these are not coordinate bases. I.e., (\ref{Gamma_mu covector})$_1$ is true also if ``non-holonomic" frame fields are chosen for TV. 
\footnote{\label{Connection matrices}\
Let $D$ be a connection on some vector bundle ${\sf E}$ with base V, let $(u_\alpha ) $ be a frame field on TV, and let $(e_a)$ be a frame field on ${\sf E}$. The connection matrices $\Gamma _\alpha $ of $D$ in the frame fields $(u_\alpha) $ and $(e_a)$ are defined by their scalar components $(\Gamma _\alpha)^b_{\ \,a} $, such that 
\be\label{De_a}
De_a(u_\alpha) =(\Gamma _\alpha)^b_{\ \,a}\, e_b. 
\ee 
This leads immediately to (\ref{Gamma_mu covector})$_1$. If the frame field on TV is a local coordinate basis: $u_\alpha =\delta ^\mu _\alpha \partial _\mu $, one may then compute the covariant derivatives $D_\mu  \Psi ^b$ of any section of ${\sf E}$, $\psi =\Psi ^b e_b$, in a matrix form: $ D_\mu \Psi =\partial _\mu \Psi + \Gamma _\mu \Psi $. Thus, this notion of a connection matrix \cite{A45} extends conveniently the usual notion of the matrices of the ``spin connection" entering the covariant Dirac equation, to any connection on a general vector bundle. It has a simple relation to the definition of a connection ``matrix" as a matrix of one-forms \cite{ChernChenLam1999}, $\omega =(\omega ^b_{\ \,a})$: if $(\theta ^\beta )$ is the dual frame of a frame field $(u_\alpha ) $ on TV, one has $\omega ^b_{\ \,a}= (\Gamma _\alpha)^b_{\ \,a}\, \theta ^\alpha $. The covector transformation of the matrices $\Gamma _\alpha $ on changing $(u_\alpha )$ applies for a given frame field on ${\sf E}$ in (\ref{De_a}), thus it does not apply if ${\sf E}=\mathrm{TV}$ and $e_a=\delta ^\alpha _a u_\alpha $.
} 
This will be useful to us because, for the DFW equation, the expression of the connection matrices (of the spin connection defined on the spinor bundle) is simple if, as the frame field on TV, one chooses precisely the tetrad field $(u_\alpha )$ used in the definition (\ref{flat-deformed}). These connection matrices are then \cite{HehlNi1990, Ryder2008}:
\be\label{Spin connection with tetrad field}
\Gamma^\sharp_\epsilon  =\frac{1}{8}\gamma _{\alpha \beta \epsilon } \,s ^{\alpha \beta }, \qquad s ^{\alpha \beta }\equiv \left[ \gamma ^{\sharp \alpha },\gamma ^{ \sharp \beta }\right].
\ee
Here $\gamma _{\alpha \beta \epsilon }\equiv \eta _{\alpha \zeta}\gamma^\zeta_{\ \beta \epsilon } $, where $\eta _{\alpha \zeta}\equiv \mathrm{diag}(1,-1,-1,-1)$ is the Minkowski ``metric" (in Cartesian coordinates) and the $\gamma^\zeta_{\ \beta \epsilon }$ 's are the coefficients of the Levi-Civita connection on TV. With an orthonormal tetrad field like $(u_\alpha )$, the $\gamma _{\alpha \beta \epsilon }$ 's can be calculated as \cite{HehlNi1990, Ryder2008}:
\be\label{gamma_alpha beta epsilon}
\gamma _{\alpha \beta \epsilon }=-\frac{1}{2}\left(C_{\alpha \beta \epsilon }+C_{\beta \epsilon \alpha }-C_{\epsilon \alpha \beta } \right)=-\gamma _{\beta \alpha \epsilon },
\ee
where $C_{\alpha \beta \epsilon }\equiv \eta _{\alpha \zeta}C^\zeta_{\ \beta \epsilon } =-C_{\alpha \epsilon \beta }$, the $C^\zeta_{\ \beta \epsilon } $ 's being the coefficients of the decomposition, in the tetrad basis, of the commutators of the same tetrad:
\be\label{structure constants}
[u_\beta,u_\epsilon ]=C^\zeta_{\ \beta \epsilon } u_\zeta .
\ee

\section{Dependences of the Hamiltonian on the reference frame and on the tetrad field}\label{Dependence}

\paragraph{The relation (\ref{purely-spatial-change})}\label{ReferenceFrame}  between two charts (coordinate systems) is an equivalence relation for charts which are all defined on a given (open) domain U of the spacetime. We call {\it reference frame} an equivalence class for this relation. Thus, if $\chi: X \mapsto (x^\mu )$ is some chart, defined on some domain U, one defines a reference frame by considering the class of $\chi $, that is, the set F of all charts which are defined on U and which exchange with $\chi $ by a purely spatial  change (\ref{purely-spatial-change}). A physically admissible reference frame is one for which we have $g_{00}>0$ everywhere in U, which condition is invariant under a change (\ref{purely-spatial-change}). The data of a physically admissible reference frame F determines \cite{A47,Cattaneo1958} a unique four-velocity vector $v=v_\mathrm{F}$: in any chart belonging to F, its components are given by
\be\label{v_F^mu}
\left(v_\mathrm{F} \right)^0\equiv \frac{1}{\sqrt{g_{00}}}, \qquad \left(v_\mathrm{F} \right)^j=0.
\ee
Note that the vector $v_\mathrm{F}$ is indeed invariant under a change (\ref{purely-spatial-change}). Equation (\ref{v_F^mu}) may be rewritten as 
\be\label{v_F}
v_\mathrm{F}=\partial _0/\sqrt{g_{00}},
\ee
with $(\partial _\mu )$ the natural basis of any coordinate system belonging to the frame F. This definition of a reference frame formalizes Cattaneo's idea of a reference fluid as a  three-dimensional congruence of time-like world lines. The world lines of the congruence have constant space coordinates, in any chart $\chi $ of F. The vector field $v_\mathrm{F}$ is the normed tangent vector field to these world lines. However, in addition, this definition fixes the time coordinate. This is necessary, because the Hamiltonian operator H does depend on the choice of the time coordinate. See Ref. \cite{A47} and references therein for more detail.\\

Thus, the invariance of $\mathrm{H}$ under the changes (\ref{purely-spatial-change}) means that H depends on the coordinate system only through the reference frame. {\it The dependence of $\mathrm{H}$ on the reference frame is natural} \cite{A48,A47,A42} and does not imply that the choice of the reference frame should be restricted in any way beyond the necessity of considering a physically admissible reference frame, i.e., one such that the world lines of the congruence are time-like. However, one needs a prescription for choosing the {\it tetrad field,} which  by Eq. (\ref{flat-deformed}) determines the coefficient field $\gamma ^\mu $ in the Dirac equation (\ref{Dirac-normal}). To see this, first note that
\bi

\item The data of a tetrad field $(u_\alpha )$ is {\it more} than the data of a reference frame, because already the time-like vector $u_0$ of the tetrad determines a congruence of world lines --- namely, the integral lines of $u_0$ \cite{A47}. 

\item Until recently, the choice of the tetrad field has been assumed to be entirely neutral, so it has been assumed that one can {\it independently} fix the reference frame and choose the tetrad field.  Thus, the tetrad field need not be ``adapted" in the sense of Eq. (\ref{u_0=v_F}) below to the (arbitrary) chosen reference frame. 
\ei
However, it was proved \cite{A43} that, if the choice of the tetrad field $(u_\alpha )$ is left free, the energy spectrum {\it in a given reference frame} --- or even in a given coordinate system --- is not unambiguously defined. This applies already to an inertial frame in a Minkowski spacetime \cite{A47}, and this also in the presence of an electromagnetic field, so that even the energy levels of the hydrogen atom would not be defined \cite{A48}. Concrete examples of the dependence of the operator H (and E, see below) on the tetrad field, for a given reference frame, will be given in this paper. {\it That} dependence is {\it not} natural. Note for example that, in contrast with the DFW Hamiltonian, the Hamiltonian associated with Dirac's original equation valid only in Cartesian coordinates in a Minkowski space {\it is} fixed once has chosen an inertial reference frame \cite{A40}.\\

The relevant Hilbert-space scalar product was derived uniquely from rather compelling conditions \cite{A42}. This scalar product involves the hermitizing matrix $A$ \cite{Pauli1936}, which for a general $(\gamma ^\mu )$ field is also a field, $A=A(X)$ \cite{A42}. However, usually the $(\gamma ^\mu )$ field is deduced from a tetrad field and from a constant set of ``flat" Dirac matrices $(\gamma ^{\sharp \alpha })$ as in Eq. (\ref{flat-deformed}). Then any hermitizing matrix for the set $(\gamma ^{\sharp \alpha })$ is also a (constant) hermitizing matrix $A$ for the $(\gamma ^\mu )$ field \cite{A42}. This is the relevant case for the present work. Moreover, in the literature, the set $(\gamma ^{\sharp \alpha })$ is usually chosen such that the hermitizing matrix is simply $A=\gamma ^{\sharp 0}$. In that particular case, the scalar product has the form proposed by Parker \cite{Parker1980} and by Leclerc \cite{Leclerc2006}. When the operator H is not Hermitian for the scalar product (which is the general case with a non-stationary metric \cite{Leclerc2006,A42,Parker1980}), one should replace H by its Hermitian part or ``energy operator" E. The latter has the physically important property that the ``field energy" $E$ associated with the Dirac field obeying Eq. (\ref{Dirac-normal}), is equal to the mean value of the energy operator E \cite{A43, Leclerc2006, A48}. However, in the present paper, only time-independent metrics and Dirac matrices $\gamma ^0 $ will occur. Therefore, the hermiticity condition proved in Ref. \cite{A42}:
\footnote{\ 
When $A$ is the constant $A=\gamma ^{\sharp 0}$, the hermiticity condition has been derived in the form $(\forall \Psi ,\Phi )\ \int \Psi ^\dagger \gamma ^{\sharp 0} \partial _0 \left(\sqrt{-g}\,  \gamma^0  \right) \Phi \, \dd ^3{\bf x}=0$ by Parker \cite{Parker1980} and by Huang \& Parker \cite{HuangParker2009}. A particular case of the latter integral condition has been derived by Leclerc \cite{Leclerc2006}.
}
\be\label{hermiticity-condition}
\partial _0 \left(\sqrt{-g}\,  A \gamma^0 \right) =0,\qquad g\equiv \mathrm{det}(g_{\mu \nu })
\ee
is verified, so that {\it in the present paper the energy operator coincides with} H.

\section{Different prescriptions for uniqueness}\label{Prescriptions}

We will now give precisions about the two different frameworks \cite{A47,A48} which we proposed in order to {\it restrict the choice of the tetrad field} consistently and sufficiently, and which were outlined in Section \ref{Intro}. The first framework involves consideration of ``spatial tensors" (e.g. ``spatial vectors"), which can be defined rigorously as tensor fields on the ``space manifold" M associated with a given \hyperref[ReferenceFrame]{reference frame} F \cite{A47}. Here we will use simple words. As we recalled, the Hamiltonian, as well as the energy operator, depend naturally on the reference frame
. This means that, to get a unique Hamiltonian operator, we need first to fix a reference frame. This can be done by considering a given physically admissible coordinate system $(x^\mu )$ defined on some domain U of the spacetime. However, we may replace the coordinate system by another one, provided this is related to the starting one by a change (\ref{purely-spatial-change}). Let us summarize successively the two different frameworks.

\paragraph{Framework I.}\label{Framework I} The data of a reference frame F fixes its four-velocity field $v_\mathrm{F}$, Eq. (\ref{v_F^mu}). Now the vector field $u_0$ of an orthonormal tetrad field $(u_\alpha )$  is time-like and normed, hence it is also a four-velocity. To attach the tetrad with the reference frame, on should thus impose the condition \cite{A47,MashhoonMuench2002,MalufFariaUlhoa2007}
\be\label{u_0=v_F}
u_0 = v_\mathrm{F}.
\ee
Let us call this an ``adapted" tetrad field to the considered reference frame F. There are many different tetrad fields which are adapted to a given arbitrary reference frame, since no condition is imposed on the vectors $u_p\ (p=1,2,3)$ beyond the orthonormality of the whole tetrad $(u_\alpha )$. However, the latter condition implies \cite{A47} that the following tensor is antisymmetric:
\be\label{Phi ST}
\Phi _{\alpha \beta } \equiv \Mat{g}\left(u_\alpha ,\left(\frac{Du_\beta }{dt }\right)_C \right) =- \Phi _{\beta \alpha },
\ee
where $\left(\frac{Du}{d\xi }\right)_C$ designates the absolute derivative, with respect to the arbitrary parameter $\xi $ along some curve $C$ in the spacetime, of a vector $u=u(\xi )$. Here specifically, for any point $X$ in the domain U, we take $C$ to be that unique world line $x(X)$ of the congruence attached to the reference frame F which passes at $X$: in any chart of F, the spatial coordinates $x^j$ are fixed along $x(X)$ and only the coordinate time $t\equiv x^0/c$ varies; $C$ is parameterized by $t$. We define thus $\Phi _{\alpha \beta }(X)$, for any point $X \in \mathrm{U}$. One shows \cite{A47} that
\be\label{Phi explicit}
\Phi _{\alpha \beta }= c \frac{d\tau }{dt}\gamma _{\alpha \beta 0},
\ee
where $\tau $ is the proper time along the world line $x(X)$ and the coefficients $\gamma _{\alpha \beta \epsilon }$ are given by Eq. (\ref{gamma_alpha beta epsilon}). Moreover, one shows that the spatial components $\Phi _{p q }\ (p,q=1,2,3)$ make a spatial tensor $\Mat{\Phi }$ in a precise geometrical sense. This tensor is indeed the opposite of the {\it rotation rate of the spatial triad} $({\bf u}_p)$. [To any four-vector $u$ --- here $u_p\ (p=1,2,3)$ --- we associate the spatial vector ${\bf u}$ --- here ${\bf u}_p$ --- whose components are the spatial components $u^j$ of $u$ in a chart belonging to the \hyperref[ReferenceFrame]{reference frame} F considered. This spatial vector is independent of the chart $\chi \in \mathrm{F}$ since, on changing the chart $\chi \in \mathrm{F}$ by (\ref{purely-spatial-change}), the components $u^j$ transform correctly.] The rotation rate of the spatial triad is also a spatial tensor $\Mat{\Xi }$, whose components in the triad basis $({\bf u}_p)$ are thus:
\be\label{Xi=-Phi}
\Xi_{p q }=-\Phi_{p q }=-c \frac{d\tau }{dt}\gamma _{p q 0}.
\ee
It has also been proved that, if two tetrad fields are adapted to the same reference frame F and if the associated spatial triads have the same rotation rate $\Mat{\Xi }$, then the two tetrad fields give rise, in that reference frame $\mathrm{F}$, to physically equivalent Dirac Hamiltonian operators, as well as to physically equivalent Dirac energy operators. Thus the first framework for uniqueness consists, in a given reference frame, in choosing an {\it adapted} tetrad field such that, in addition, its rotation rate tensor field $\Mat{\Xi }$ is a predefined field. Two natural choices are:
\bi 
\item {\it a}) $\Mat{\Xi }=\Mat{\Omega }$, where $\Mat{\Omega }$ is the rotation-rate field of the reference frame F itself \cite{A47,Cattaneo1958,Weyssenhoff1937}, whose components in a coordinate system of F are
\footnote{\label{Omega vs t}\
The spatial tensor $\Mat{\Omega }$ depends on the choice of the time coordinate $t$ in a complex manner, whereas, on changing from $t$ to $t'$, \ $\Mat{\Xi }$ gets simply multiplied by $dt/dt'$. Hence, the equality $\Mat{\Xi }=\Mat{\Omega  }$ is not covariant under a change of the time coordinate, so that the prescriptions $\Mat{\Xi }=\Mat{\Omega  }$ corresponding to reference frames differing merely in the choice of the time coordinate are not physically equivalent. And indeed, there is a rewriting of the geodesic equation of motion in the form of Newton's second law, in which the tensor $\Mat{\Omega }$ plays exactly the role played by the angular velocity tensor of a rotating frame in Newtonian theory \cite{Cattaneo1958} --- but in this rewriting $\Mat{\Omega }$ has to be calculated with a time coordinate $\hat{x}^0$ such that, along a world line of the congruence, we have $d\hat{x}^0=c\,d\tau $, where $d\tau $ is the proper time increment. Thus, if one applies the prescription $\Mat{\Xi }=\Mat{\Omega}$, one should impose that the time coordinate be a such one, $\hat{x}^0$ with $d\hat{x}^0=c\,d\tau $ \cite{A47}. For the uniformly rotating frame, the tensors $\Mat{\Omega  }$ calculated with either $t$ or $\tau $ differ only by $O(V^2/c^2)$ \cite{A47} ($V$ is defined in Sect. \ref{inertial/rotating}), and the same is easy to check for $\Mat{\Xi }$.
}
\be\label{Weyssenhoff modified}
\Omega_{jk} \equiv \frac{1}{2}c\sqrt{g_{00}}\,(\partial _j g_k-\partial _k g_j-g_j\partial _0 g_k+g_k\partial _0 g_j),\qquad g_j\equiv \frac{g_{0j}}{g_{00}}. 
\ee

\item {\it b}) $\Mat{\Xi }=\Mat{0 }$.
\ei
As we announced in the Introduction, the two choices {\it a}) and {\it b}) lead to non-equivalent Hamiltonians, thus represent two different solutions to the non-uniqueness problem \cite{A47}.\\

That first framework is difficult to implement, especially its variant {\it a}) which needs to calculate the field $\Mat{\Omega }$ and to find a tetrad field such that $\Mat{\Xi }=\Mat{\Omega}$: in practice, this could be done only approximately, by numerical integration of ordinary differential equations of the form $\delta {\bf u}_q /dt  = \Omega^p _{\ \,q }\, {\bf u}_p$. \{Here $\delta {\bf u}_q /dt $ is  the Fermi-Walker derivative of ${\bf u}_q$ \cite{A47}.\} Moreover, by imposing the condition (\ref{u_0=v_F}), we limit from the outset the validity of this kind of solution of the non-uniqueness problem to a given reference frame.

\paragraph{Framework II.}\label{Framework II} That framework needs that there is some special coordinate system $(x^\mu )$, in which the metric has the special form (\ref{isotropic-diagonal}) \cite{A48}. As discussed there, this form is general enough for the prospective purpose of testing the generally-covariant Dirac equations in a realistic spacetime metric. 
\footnote{\
Moreover, this form is generic for an alternative theory of gravitation  \cite{A35}, in the preferred reference frame assumed by that theory. That theory is based only on a scalar field which determines, among other things, the physical metric $\Mat{g}$, from an a priori assumed flat metric, say $\Mat{\gamma }$. Although it thus has two metrics, this is not a metric theory in the standard sense.
}
Then one chooses the ``diagonal tetrad" in that coordinate system, Eq. (\ref{diagonal-tetrad}). This defines the Dirac matrices $\gamma ^\mu $ in that coordinate system, Eq. (\ref{flat-deformed}). Then, in any possible coordinate system, say $(x'^\mu )$, the Dirac matrices $\gamma '^\mu $ are  got by the transformation (\ref{gamma^mu vector}). As it has been proved in Ref. \cite{A48}: if one considers another coordinate system in which the metric has also the {\it form} (\ref{isotropic-diagonal}) (a priori not with the same coefficients), then one passes from the first to the second one by a constant rotation, combined with a constant homothecy. It follows \cite{A48}, first, that the corresponding ``diagonal tetrads" (\ref{diagonal-tetrad}) exchange by a {\it constant} Lorentz transformation, and then, that in any given reference frame, the Hamiltonian operators got from these two choices of tetrad fields are equivalent, as well as the energy operators. Thus the non-uniqueness problem is solved simultaneously in any possible reference frame, and in a simple tractable way.

\section{Dirac energy operator in an inertial or a rotating frame}\label{inertial/rotating}

Starting with a global inertial reference frame F$'$ in a Minkowski spacetime, defined from a Cartesian system of coordinates $(x'^\mu )=(ct',x',y',z')$, we define the uniformly rotating reference frame F from the rotating coordinates $(x^\mu)=(ct,x,y,z)$ given by
\be\label{rotating Cartesian}
t=t',\quad x=x'\cos \omega t + y' \sin \omega t,\quad y=-x' \sin \omega t + y' \cos \omega t,\quad z=z',
\ee
where $\omega $ is a real constant. In these coordinates, the Minkowski metric remains stationary: it becomes 
\be\label{Minkowski in rotating Cartesian}
ds^2=\left[1-\left(\frac{\omega }{c}\right)^2(x^2+y^2)\right](dx^0)^2 + 2\frac{\omega}{c} (y\,dx-x\, dy)\, dx^0-(dx^2+dy^2 +dz^2).
\ee 
The validity of these new coordinates is restricted by the admissibility condition $g_{00}>0$ to the domain $\mathrm{U}$ made of those points in the spacetime for which we have $\ V\equiv \omega \rho <c$, where $\rho \equiv (x^2+y^2)^{1/2}$. Thus, in contrast with the inertial frame F$'$, the rotating frame F is a local reference frame, so that going from F$'$ to F represents some ``loss of information". Indeed the Hamiltonian and energy operators in the frame F act on wave functions $\Psi $ defined on the spatial manifold M associated with F \cite{A43}. The extension of that manifold depends on the domain $\mathrm{U}$ of the coordinates considered \cite{A44}. That is, the operators H and E act on wave functions $\Psi $ depending on the spatial coordinates $x,y,z$, whose domain of definition is only a subset U of the whole spacetime --- specifically, here U is defined by the condition $\omega \rho <c$. However, if the rotating frame follows the rotation of a real astronomical object, this limitation does not have any practical consequence. Even for the extreme case of a neutron star with the highest observed angular velocity $\omega \simeq 10^3/\mathrm{s}$, the limitation only imposes $\rho < 3.10^5\,\mathrm{m}$, which is still 30 times the typical radius of the neutron star, $R\simeq 10\, \mathrm{km}$. Clearly, at such distances the wave function of, say, a neutron, can safely be equated to zero. For the Earth, with $\omega \simeq 7.10^{-5}/\mathrm{s}$, the limitation is $\rho <5.10^{12}\,\mathrm{m}\simeq 30\,\mathrm{au}$. \\


\subsection{Energy operators in the two frames with the Cartesian tetrad}

In the global Cartesian coordinates $(x'^\mu )$ on the Minkowski spacetime, the metric has of course the space-isotropic diagonal form (\ref{isotropic-diagonal}), hence we can apply \hyperref[Framework II]{Framework II}. The corresponding diagonal tetrad (\ref{diagonal-tetrad}) is just $u'_\alpha \equiv \delta ^\mu _\alpha \partial '_\mu $, that is, the natural basis of the Cartesian coordinate system $(x'^\mu )$, or ``Cartesian tetrad". Clearly, the coefficients $\gamma^\zeta_{\ \beta \epsilon }$ of the Levi-Civita connection are zero with this tetrad field $(u'_\alpha )$, 
\footnote{\
Hence, by (\ref{Xi=-Phi}): for that tetrad, in the inertial frame F$'$ to which it is adapted, we have $\Mat{\Xi}=\Mat{0}$: {\it the Cartesian tetrad solves Variant {\it b}) of \hyperref[Framework I]{Framework I} for the inertial frame.}
} 
so the connection matrices (\ref{Spin connection with tetrad field}) are $\Gamma ^\sharp _\epsilon =0$ and, by (\ref{Gamma_mu covector}), they remain zero in any coordinates. Also, using the tetrad $(\partial '_\mu )$, the Dirac matrices (\ref{flat-deformed}) in the coordinates $(x'^\mu )$ are simply the ``flat" matrices, $\gamma '^\mu =\gamma ^{\sharp \mu }$. Hence, when it is used in the inertial frame F$'$ itself, the Cartesian tetrad yields by (\ref{Hamilton-Dirac-normal}) just the special-relativistic Hamiltonian, which is Hermitian:
\be \label{Hamilton-Dirac-SR}
\mathrm{E}'_1 = \mathrm{H}'_1 =  mc^2\gamma ^{\sharp 0} -i\hbar c\alpha ^{\sharp j} \partial'_j,
\ee
where $\alpha ^{\sharp j} \equiv \gamma ^{\sharp 0}\gamma  ^{\sharp j}$. This result is the physically correct one: in an inertial frame, the Hamiltonian operator should indeed be the one predicted by Dirac's original theory.\\

The Hamiltonian H$_1$ in the rotating frame F and corresponding with the Cartesian tetrad involves the Dirac matrices transformed to the rotating coordinates (\ref{rotating Cartesian}) by Eq. (\ref{gamma^mu vector}):
\bea\label{gamma rotating}
\gamma ^0 & = & \gamma ^{\sharp 0},\quad \gamma ^1=\gamma ^{\sharp 1}\cos \omega t +  \gamma ^{\sharp 2}\sin \omega t+\frac{\omega y}{c}\,\gamma ^{\sharp 0},\\
\gamma ^2 & = & -\gamma ^{\sharp 1}\sin \omega t +\gamma ^{\sharp 2}\cos \omega t -\frac{\omega x}{c} \,\gamma ^{\sharp 0},\quad \gamma ^3=\gamma ^{\sharp 3}.
\eea
From (\ref{hermiticity-condition}), (\ref{Minkowski in rotating Cartesian}) and (\ref{gamma rotating})$_1$, it follows that H$_1$ is Hermitian. Noting that $g^{00}=1$ after the coordinate change (\ref{rotating Cartesian}), we get then the $\alpha $ matrices of Eq. (\ref{alpha}):
\bea\label{alpha-Minkowski-tetrad-1}
\alpha ^0 & = & \gamma ^{\sharp 0},\quad \quad \alpha ^1=\alpha^{\sharp 1}\cos \omega t +  \alpha^{\sharp 2}\sin \omega t+\frac{\omega y}{c}\,{\bf 1}_4,\\
\label{alpha-Minkowski-tetrad-2}
\alpha  ^2 & = & -\alpha ^{\sharp 1}\sin \omega t +\alpha ^{\sharp 2}\cos \omega t -\frac{\omega x}{c}\,{\bf 1}_4,\quad \alpha ^3=\alpha ^{\sharp 3}.
\eea
We have moreover from (\ref{rotating Cartesian}):
\be\label{d'_j fn d_k}
\cos \omega t \, \partial _x-\sin \omega t\, \partial _y=\partial _{x'},\quad \sin \omega t\, \partial _x + \cos \omega t \, \partial _y=\partial _{y'},\quad \partial_z=\partial _{z'}.
\ee
Therefore, the energy operator $\mathrm{E}_1 = \mathrm{H}_1$, Eq. (\ref{Hamilton-Dirac-normal}), is explicitly:
\bea
\mathrm{H}_1 & = & mc^2\alpha ^0 -i\hbar c \alpha ^{ j} \partial_j \nonumber\\
& = & mc^2\gamma ^{\sharp 0} -i\hbar c \left[\alpha ^{\sharp 1}(\cos \omega t \, \partial _x-\sin \omega t\,\partial _y)+\alpha ^{\sharp 2}(\sin \omega t\, \partial _x + \cos \omega t \, \partial _y)+\alpha ^{\sharp 3} \partial _z \right]\nonumber \\
& & -i\hbar c \frac{\omega}{c} (y\partial _x-x\partial _y){\bf 1}_4\nonumber \\
& = & \mathrm{H}'_1 - i\hbar \omega (y\partial _x-x\partial _y),\nonumber\\
\mathrm{H}_1 & = & \mathrm{H}'_1 -\Mat{\omega .}{\bf L}. \label{Hamilton-restricted-gauge}
\eea
Here, ${\bf L}\equiv {\bf r}\wedge (-i\hbar \nabla )$ is the angular momentum operator. Thus, in the case of a uniformly rotating frame in a Minkowski spacetime [and arguably in general, see Eq. (\ref{delta H-relative rotation}) below], \hyperref[Framework II]{Framework II} does not predict any spin-rotation coupling. 

\subsection{Constructing a tetrad adapted to the rotating frame}

Let us now try to use \hyperref[Framework I]{Framework I}. One rotating orthonormal tetrad that appears naturally in the metric (\ref{Minkowski in rotating Cartesian}) is Ryder's \cite{Ryder2008} first tetrad:
\be\label{Ryder1}
u_0=\frac{1}{c}\frac{\partial }{\partial t}+\frac{\omega y}{c}\frac{\partial }{\partial x}-\frac{\omega x}{c}\frac{\partial }{\partial y}, \quad u_1=\frac{\partial }{\partial x}, \quad u_2=\frac{\partial }{\partial y}, \quad u_3=\frac{\partial }{\partial z}.
\ee
However, as noted in Ref. \cite{A47}, it results from (\ref{rotating Cartesian}) and (\ref{Ryder1}) that
\bea\label{Ryder 1 = Cartesian tetrad-0}
u_0 & = & \frac{\partial x^\nu }{\partial x'^0 }\frac{\partial }{\partial x^\nu }=\frac{\partial }{\partial x'^0 }\equiv \partial'_0,
\\
\nonumber\\
\label{Ryder 1 = Cartesian tetrad-123}
u_1 & = & \cos \omega t \,\partial '_1+\sin \omega t\,\partial'_2,\quad u_2=-\sin \omega t \,\partial'_1+\cos \omega t\,\partial '_2, \quad u_3=\partial'_3,
\eea
where $(\partial '_\mu )$ is the Cartesian tetrad. Since $v_\mathrm{F}=\partial _0/\sqrt{g_{00}}$, Eq. (\ref{v_F}), Equation (\ref{Ryder 1 = Cartesian tetrad-0}) means that Ryder's tetrad $(u_\alpha )$ is ``adapted" in the sense of Eq. (\ref{u_0=v_F}) to the inertial frame F$'$, not to the rotating frame F.  On the other hand, since $\Mat{g}(\partial _\mu ,\partial _\nu )=g_{\mu \nu }$, we see from (\ref{Minkowski in rotating Cartesian}) that the natural basis $(\partial _\mu )$ of the rotating coordinates $x^\mu $ given by (\ref{rotating Cartesian}) is not orthogonal. But consider, at each $X\in \mathrm{U}$, the hyperplane $\mathrm{H}_X$ in the local tangent space $\mathrm{TV}_X$ to the spacetime V, made of the vectors which are orthogonal to $v_\mathrm{F}(X)$. Define the orthogonal projection $\Pi_X$ onto $\mathrm{H}_X$ \cite{A47,JantzenCariniBini1992}. (Note that this operator depends on the reference frame F which is considered, as do $v_\mathrm{F}$ and $\mathrm{H}_X$.) Obviously, if at each $X\in \mathrm{U}$ we thus project the vectors $\partial _j(X)\ (j=1,2,3)$ onto $\mathrm{H}_X$, we get vector fields $\Pi  \partial _j$ such that $(\Pi  \partial _j)(X)$ is orthogonal to $v_\mathrm{F}(X)$ at any $X\in \mathrm{U}$. From the definition, one finds the components of $\Pi _X a$ for a vector $a \in \mathrm{TV}_X$, in a coordinate system belonging to F \cite{A47}. Thus we get
\be\label{Pi_d_j}
(\Pi  \partial _j)^0=-g_{0k}(\partial _j)^k/g_{00}=-g_{0j}/g_{00},\quad (\Pi  \partial _j)^k=(\partial _j)^k = \delta ^k_j,
\ee 
from which it follows that 
\be\label{g(Pi d_j, Pi d_k)}
\Mat{g}(\Pi \partial _j,\Pi \partial _k)=g_{jk}-\frac{g_{0j}g_{0k}}{g_{00}} \equiv  -h_{jk}.
\ee
Here $\Mat{h}$ is the spatial metric of the reference frame F, such that for any two vectors $a,b$ at $X$ \cite{A47,JantzenCariniBini1992}:
\be\label{Def h}
\Mat{h}_X(a,b)  \equiv  -\Mat{g}_X(\Pi _X a,\Pi _X b) .
\ee
Note that the definition of $\mathrm{H}_X$ and $\Pi_X$, as well as Eqs. (\ref{Pi_d_j}) to (\ref{Def h}), are valid for a general reference frame in a general spacetime. Coming back to the uniformly rotating frame in a Minkowski spacetime, from (\ref{Minkowski in rotating Cartesian}) and (\ref{g(Pi d_j, Pi d_k)}) we get that $\Mat{g}(\Pi \partial _1,\Pi \partial _3)=\Mat{g}(\Pi \partial _2,\Pi \partial _3)=0$ but 
$\Mat{g}(\Pi \partial _1,\Pi \partial _2) \ne 0$, $(\partial _\mu )$ being again specifically the natural basis associated with the coordinates (\ref{rotating Cartesian}). Hence, one may define an orthonormal tetrad adapted to the rotating frame F by taking: $v_\mathrm{F}$, $\Pi \partial _2/\parallel \Pi \partial _2\parallel $, $\Pi \partial _3/\parallel \Pi \partial _3\parallel $, and the vector product of (the spatial vectors associated with) the two last vectors. However, a simpler orthonormal tetrad adapted to F is obtained by considering the natural basis $(\partial ^\circ_\mu )$ of the ``rotating cylindrical coordinates" $(x^{\circ \mu })=(ct, \rho , \varphi , z)$, related to the coordinates (\ref{rotating Cartesian}) by  
\be\label{rotating cylindrical}
x=\rho \cos\varphi,\quad y=\rho  \sin\varphi .
\ee
(It follows from this that $\partial ^\circ_0=\partial _0 $ and $\partial ^\circ_3=\partial ^\circ_z=\partial _3=\partial _z $.) In the coordinates $(x^{\circ \mu })$, the Minkowski metric (\ref{Minkowski in rotating Cartesian}) rewrites immediately as
\be\label{L&L(89,2)}
ds^2=g^\circ_{\mu \nu }dx^{\circ \mu }dx^{\circ \nu }=\left[1-\left(\frac{\omega \rho }{c}\right)^2\right]c^2 dt^2 - 2\omega \rho ^2 \,d\varphi\, dt-(d\rho ^2+\rho ^2 d\varphi ^2+dz^2),
\ee
from which we find that the spatial metric defined by Eq. (\ref{g(Pi d_j, Pi d_k)})$_2$ has components [in the coordinates $(x^{\circ j})=(\rho ,\varphi ,z)$]:
\be\label{h-cylindrical}
h_{jk}=\delta _{jk} \quad \mathrm{except\ for}\ \ h_{22}=\frac{\rho ^2}{1-\omega ^2 \rho ^2/c^2}.
\ee
Hence, owing to Eq.  (\ref{g(Pi d_j, Pi d_k)})$_1$, we define an orthonormal tetrad adapted to the rotating frame F by taking $v_\mathrm{F}=\partial ^\circ_0/\sqrt{g^\circ_{00}}$ and by norming the $\Pi\partial ^\circ_j $ vectors, which results simply in setting
\be\label{u circ-01}
u^\circ_0=\frac{1}{\sqrt{1-\omega ^2 \rho ^2/c^2}}\partial _0,\qquad u^\circ_1 = \partial ^\circ_1 =\partial ^\circ_\rho ,
\ee
\be\label{u circ-23}
\quad u^\circ_2 = \frac{\omega \rho }{c\sqrt{1-\omega ^2 \rho ^2/c^2}}\partial _0+\frac{\sqrt{1-\omega ^2 \rho ^2/c^2}}{\rho }\partial^\circ _\varphi ,\qquad u^\circ_3 =\partial ^\circ_3 = \partial ^\circ_z =\partial _z.
\ee
We note that the matrix $a\equiv (a^\mu_{\ \,\alpha})$, such that $u^\circ_\alpha =a^\mu_{\ \,\alpha}\,\partial^\circ_\mu$, is independent of the time coordinate $t$. Hence, so are also the Dirac matrices (\ref{flat-deformed}). Thus, from (\ref{hermiticity-condition}), the Hamiltonian operator in the rotating frame with the adapted rotating tetrad is Hermitian. \\

Let us calculate the rotation rate tensor field $\Mat{\Xi }$ of the tetrad $(u^\circ_\alpha)$, Eq. (\ref{Xi=-Phi}). The coefficients of the decomposition (\ref{structure constants}) of the commutators of the tetrad (\ref{u circ-01})-(\ref{u circ-23}) are easily computed to be: $C^\zeta _{\ \,\beta \epsilon}=0$, except for:
\be\label{C^zeta _beta epsilon-1}
 C^0 _{\ \,0 1}=-C^0 _{\ \,1 0}=-\frac{\rho \omega ^2}{c^2-\omega ^2 \rho ^2},\quad C^0 _{\ \,1 2}=-C^0 _{\ \,2 1}=\frac{2 \omega }{c(1-\omega ^2 \rho ^2/c^2)},
\ee
\be\label{C^zeta _beta epsilon-2}
 C^2 _{\ \,1 2}=-C^2 _{\ \,2 1}=-\frac{1}{\rho (1-\omega ^2 \rho ^2/c^2)}.
\ee
From this, we deduce immediately the coefficients $C_{\alpha \beta \epsilon }= \eta _{\alpha \zeta}C^\zeta_{\ \beta \epsilon } $, then we get the coefficients $\gamma _{\alpha \beta \epsilon }=-\gamma _{ \beta \alpha \epsilon }$ [Eq. (\ref{gamma_alpha beta epsilon})]. They are zero, except for (when $\alpha < \beta$):
\be\label{gamma_alpha beta epsilon-1}
 \gamma_{0 1 0}=-\frac{\rho \omega ^2}{c^2-\omega ^2 \rho ^2},\quad \gamma _{1 2 2}=\frac{1 }{\rho (1-\omega ^2 \rho ^2/c^2)},
\ee
\be\label{gamma_alpha beta epsilon-2}
 \gamma_{1 2 0}=-\gamma_{0 1 2}=\gamma_{0 2 1}=\frac{ \omega}{c(1-\omega ^2 \rho ^2/c^2)}.
\ee
Therefore, Eqs. (\ref{Xi=-Phi}) and (\ref{L&L(89,2)}) give us: $\ \Couleur{\Xi _{p q}=0}$, except for 
\be\label{Xi rotating frame}
\Xi _{21}=-\Xi _{12}=\omega \gamma_\mathrm{L}, \qquad \gamma_\mathrm{L} =\gamma_\mathrm{L} (\rho )\equiv \left[1-(\omega^2 \rho ^2/c^2)\right]^{-1/2}.
\ee
We may compare this with the rotation rate tensor $\Mat{\Omega }$ of the reference frame, defined in general by Eq. (\ref{Weyssenhoff modified}). For the rotating frame F, the components $\Omega _{jk}$ of $\Mat{\Omega }$ are easily computed \cite{A47}:
\be\label{Omega uniformly rotating frame}
\Omega_{32}=0,\quad  \Omega_{13}=0,\quad \Omega_{21} = +\omega\gamma_\mathrm{L} ^3.
\ee
These are in fact the components of the spatial tensor $\Mat{\Omega }$ in the natural basis $(\Mat{\partial }_j)$ associated with the spatial coordinates $(x^j )$ [the spatial part of the coordinates (\ref{rotating Cartesian})]. The components $\Omega^\circ_{p q}$ of $\Mat{\Omega }$ in the spatial triad basis $({\bf u}^\circ _p )$ associated with the tetrad basis $(u^\circ _\alpha )$ are got from (\ref{Omega uniformly rotating frame}) and from the relation between the triad bases $(\Mat{\partial }_j)$ and $({\bf u}^\circ _p )$. This relation follows from (\ref{rotating cylindrical}) and (\ref{u circ-01})--(\ref{u circ-23}) and is:
\be
{\bf u}^\circ _1= \cos \varphi \,\Mat{\partial }_1+ \sin \varphi \, \Mat{\partial }_2, \quad {\bf u}^\circ _2= \left(-\sin \varphi \,\Mat{\partial }_1+ \cos \varphi \, \Mat{\partial }_2 \right)/\gamma_\mathrm{L}(\rho ),\quad {\bf u}^\circ _3= \Mat{\partial }_3.
\ee
By standard tensor transformation, we find from this and from (\ref{Omega uniformly rotating frame}):
\be\label{Omega uniformly rotating frame-2}
\Omega^\circ_{32}=0,\quad  \Omega^\circ_{13}=0,\quad \Omega^\circ_{21} = \Omega_{21}/\gamma_\mathrm{L} =\omega \gamma_\mathrm{L}^2.
\ee
These differ from (\ref{Xi rotating frame}) only by $O(V^2/c^2)$ terms (for $V\equiv \omega \rho \ll c$). Up to this negligible difference, we may thus consider that the adapted rotating tetrad $(u^\circ _\alpha )$ verifies $\Mat{\Xi }=\Mat{\Omega }$, as required by the variant {\it a}) of \hyperref[Framework I]{Framework I}.

\subsection{Energy operator with the adapted rotating tetrad}

Let us thus calculate the Hamiltonian (\ref{Hamilton-Dirac-normal}) in the rotating frame F, when choosing the tetrad $(u^\circ_\alpha )$, Eqs. (\ref{u circ-01})-(\ref{u circ-23}). We begin with the spin connection matrices (\ref{Spin connection with tetrad field}) with the tetrad field $(u^\circ_\alpha )$. From Eqs. (\ref{gamma_alpha beta epsilon-1})--(\ref{gamma_alpha beta epsilon-2}), these are:
\be\label{Gamma dièse-1}
4\Gamma ^\sharp _0=\gamma_{0 1 0}\,s ^{01}+\gamma _{1 2 0}\,s ^{1 2},\quad 4\Gamma ^\sharp _1=\gamma _{0 2 1}\,s ^{0 2},
\ee
\be\label{Gamma dièse-2}
4\Gamma ^\sharp _2=\gamma_{0 1 2}\,s ^{01}+\gamma _{1 2 2}\,s ^{1 2},\quad \Gamma ^\sharp _3=0.
\ee
To compute the connection matrices $\Gamma _\mu $ when the coordinate basis $(\partial^\circ_\mu)$ is chosen, we use the fact that they transform as a covector [see Eq. (\ref{Gamma_mu covector}) and thereafter]. Thus we have $\Gamma _\mu =b^\alpha _{\ \,\mu }\Gamma ^\sharp _\alpha $, where the matrix $b\equiv (b^\alpha _{\ \,\mu })$, such that $\partial^\circ_\mu=b^\alpha _{\ \,\mu }\,u^\circ_\alpha$, is got easily from Eqs. (\ref{u circ-01})-(\ref{u circ-23}):
\be
b=\begin{pmatrix} 
\gamma_\mathrm{L} ^{-1} & 0  & -\frac{\gamma_\mathrm{L} \omega \rho ^2}{c } & 0\\
0  & 1 & 0 & 0\\
0 & 0 & \rho \gamma_\mathrm{L} & 0\\
0 & 0  & 0 & 1
\end{pmatrix}.
\ee
We get thus from (\ref{Gamma dièse-1})--(\ref{Gamma dièse-2}), using the standard set of Dirac matrices:
\footnote{\
The choice of the set $(\gamma ^{\sharp \alpha })$ does not matter, because corresponding $(\gamma ^\mu )$ fields exchange by constant similarity transformations, hence give rise to equivalent energy operators. With the standard set (Dirac's), we have $s^{jk}=-2i\epsilon _{jkl}\Sigma ^l$ and $s^{0j}=2\Sigma'^j$.
}
\be\label{Gamma-rotating-tetrad-1}
\Gamma _0= -\frac{\gamma_\mathrm{L} }{2}\left(\frac{i \omega}{c}\, \Sigma ^3+\frac{\rho \omega^2}{c^2}\Sigma '^1 \right), \qquad \Gamma _1= \frac{i\omega \gamma_\mathrm{L} ^2}{2c}\,\Sigma'^2,
\ee
\be\label{Gamma-rotating-tetrad-2}
\Gamma _2= \frac{\gamma_\mathrm{L} ^3}{2}\left[\left(\frac{\rho \omega}{c}\right)^2-1\right] \left(\frac{\rho \omega}{c}\Sigma'^1 + i\Sigma^3 \right), \qquad \Gamma _3=0,
\ee
where
\be
\Sigma ^j\equiv 
\begin{pmatrix} 
\sigma ^j & 0  \\
0  & \sigma ^j \\
\end{pmatrix},\quad \Sigma'^j\equiv 
\begin{pmatrix} 
0 &  \sigma ^j \\
 \sigma ^j & 0\\
\end{pmatrix}.
\ee
On the other hand, the $\gamma ^\mu $ matrices are defined by (\ref{flat-deformed}). In view of Eqs. (\ref{u circ-01})-(\ref{u circ-23}), we have:
\be
\gamma ^0=\gamma_\mathrm{L}\gamma ^{\sharp 0}+ \frac{\rho \gamma_\mathrm{L}\omega }{c}\gamma ^{\sharp 2},\quad \gamma ^1=\gamma ^{\sharp 1}, \quad \gamma ^2=\frac{1 }{\rho\gamma_\mathrm{L} }\gamma ^{\sharp 2},\quad \gamma ^3=\gamma ^{\sharp 3},
\ee
from which we get the matrices $\alpha ^\mu $ of Eq. (\ref{alpha}) [note that $g^{00}=1$]:
\be\label{alpha-rotating-tetrad-1}
\alpha ^0 =\gamma ^0,\quad \alpha ^2 = \frac{1}{\rho }\left(\alpha ^{\sharp 2}+ \frac{\rho \omega }{c} {\bf 1}_4\right)=\frac{1}{\rho }\left(\Sigma'^{2}+ \frac{\rho \omega }{c} {\bf 1}_4\right),
\ee
\be\label{alpha-rotating-tetrad-2}
\alpha ^j= \gamma_\mathrm{L}\left(\alpha ^{\sharp j}+ \frac{\rho \omega }{2c} s ^{2j}\right)=\gamma_\mathrm{L}\left(\Sigma'^{j}- i\frac{\rho \omega }{c} \epsilon _{2jk}\Sigma  ^{k}\right) \quad (j=1,3).
\ee
The energy operator with the adapted rotating tetrad is thus [Eq. (\ref{Hamilton-Dirac-normal})]:
\be \label{Hamilton-Dirac-normal-2}
\mathrm{E}_2 =\mathrm{H}_2 =  mc^2\alpha  ^0 -i\hbar c\alpha ^j (\partial^\circ _j+\Gamma  _j) -i\hbar c\Gamma _0,
\ee
where the matrices $\Gamma _\mu $ and $\alpha ^\mu $ are given by Eqs. (\ref{Gamma-rotating-tetrad-1})-(\ref{Gamma-rotating-tetrad-2}) and (\ref{alpha-rotating-tetrad-1})-(\ref{alpha-rotating-tetrad-2}). In particular, for $V\equiv \rho \omega \ll c$, we have from (\ref{Gamma-rotating-tetrad-1}):
\be\label{Spin-rotation-2}
-i\hbar c\Gamma _0=-\frac{\hbar \gamma_\mathrm{L}\omega }{2}\Sigma ^3 \left [1+O\left(\frac{V}{c}\right)\right]=-\gamma_\mathrm{L}\Mat{\omega}.{\bf S}\left [1+O\left(\frac{V}{c}\right)\right]. 
\ee
which is the usual {\it ``spin-rotation coupling" term} \cite{Mashhoon1988,HehlNi1990,CaiPapini1991,Ryder2008}.

\subsection{Energy operator in the two frames with Ryder's rotating tetrad}

Since Ryder's \cite{Ryder2008} first tetrad $(u_\alpha )$, Eq. (\ref{Ryder1}) above,
is ``adapted" in the sense of Eq. (\ref{u_0=v_F}) to the inertial frame F$'$, it is interesting to compute the energy operator associated in the inertial frame F$'$ with this tetrad. We checked that, as was found by Ryder, the spin connection matrices (\ref{Spin connection with tetrad field}) for this tetrad field $(u_\alpha )$ are
\be\label{Spin connec - Ryder1}
\Gamma^\sharp_0= -\frac{i\omega}{2c}\, \Sigma ^3, \quad \Gamma^\sharp_j=0.
\ee
The tetrad $(u_\alpha )$ is related to the natural basis $(\partial '_\mu )$ by Eqs. (\ref{Ryder 1 = Cartesian tetrad-0})--(\ref{Ryder 1 = Cartesian tetrad-123}). We thus transform immediately the connection matrices to the natural basis, getting the same:
\be\label{Spin connec - Ryder1 - Cartesian}
\Gamma_0= -\frac{i\omega}{2c}\, \Sigma ^3, \quad \Gamma_j=0.
\ee
We get also from (\ref{flat-deformed}) and (\ref{Ryder 1 = Cartesian tetrad-0})--(\ref{Ryder 1 = Cartesian tetrad-123}), using then (\ref{alpha}):
\be\label{alpha - Ryder1 - Cartesian}
\alpha ^0=\gamma ^{\sharp 0},\ \alpha ^1=\cos \omega t\, \alpha  ^{\sharp 1} -\sin \omega t \,\alpha  ^{\sharp 2},\ \alpha ^2=\sin \omega t \,\alpha  ^{\sharp 1} + \cos \omega t\,\alpha  ^{\sharp 2},\ \alpha ^3=\alpha  ^{\sharp 3}.
\ee
We note that here again $\gamma ^0=\alpha ^0=\gamma ^{\sharp 0}$ is constant, so the Hamiltonian is Hermitian, Eq. (\ref{hermiticity-condition}). From (\ref{d'_j fn d_k}), (\ref{Spin connec - Ryder1 - Cartesian}), and (\ref{alpha - Ryder1 - Cartesian}), we find the 
explicit expression of the energy operator $\mathrm{E}'_3 = \mathrm{H}'_3$, Eq. (\ref{Hamilton-Dirac-normal}):
\bea 
\mathrm{H}'_3 & = & mc^2\alpha ^0 -i\hbar c \alpha ^{ j} \partial'_j -i\hbar c \Gamma _0\nonumber\\
& = & mc^2\gamma ^{\sharp 0} -i\hbar c \left[\alpha ^{\sharp 1}(\cos \omega t \, \partial _x'+\sin \omega t\,\partial _y')+\alpha ^{\sharp 2}(-\sin \omega t\, \partial _x' + \cos \omega t \, \partial _y')+\alpha ^{\sharp 3} \partial _z' \right]\nonumber \\
& & -i\hbar c \Gamma _0\nonumber, 
\eea
thus
\be
\mathrm{H}'_3 = mc^2\gamma ^{\sharp 0}  -i\hbar c \alpha ^{\sharp j} \partial_j -\frac{\hbar \omega }{2}\Sigma ^3 .
\label{Hamilton-Ryder1-inertial}
\ee
[Recall that $(\partial _\mu) $ is the natural basis of the {\it rotating} coordinates.] Thus, with Ryder's tetrad, we find that the DFW energy operator in the {\it inertial} frame F$'$ does contain the spin-rotation coupling term $-\frac{\hbar \omega }{2}\Sigma ^3=-\Mat{\omega }.{\bf S }$. This is certainly unexpected physically. Also, by comparing H$'_1$ with H$'_3$ [Eqs. (\ref{Hamilton-Dirac-SR}) and (\ref{Hamilton-Ryder1-inertial})], we have a clear confirmation of the non-uniqueness of the DFW Hamiltonian and energy operator. The energy operators H$'_1$ and H$'_3$, which are related together by a simple local similarity transformation $S$, were known in advance to be physically inequivalent \cite{A47}. They are in fact {\it grossly} inequivalent, e.g. the difference in their mean values for corresponding states $\Psi$ and $\widetilde{\Psi}\equiv S^{-1}\Psi$ depends on the state $\Psi$ and contains the {\it arbitrary} factor $\omega $. That is, for any state $\Psi=(\Psi ^\alpha )_{\alpha =0,...,3}\,$, for which the energy mean value with H$'_1$ is $\langle \mathrm{H}'_1 \rangle\equiv (\Psi \mid \mathrm{H}'_1\Psi )$, the corresponding energy mean value $\langle \mathrm{H}'_3 \rangle=(\widetilde{\Psi}\,\widetilde{\mid }\, \mathrm{H}'_3 \widetilde{\Psi})$, got by using H$'_3$, may differ arbitrarily from $\langle \mathrm{H}'_1 \rangle$ --- depending on the arbitrary rotation rate $\omega $ of Ryder's tetrad: 
\be\label{bar A-explicit}
A\equiv \langle \mathrm{H}'_3 \rangle - \langle \mathrm{H}'_1 \rangle =-\frac{\omega}{2}\int \left(\abs{\Psi^0}^2 +\abs{\Psi^2}^2-\abs{\Psi^1}^2-\abs{\Psi^3}^2\right)\,\dd^3{\bf x}
\ee 
\{Eq. (29) in Ref. \cite{A50}.\} Recall that H$'_1$ is the standard Dirac Hamiltonian of special relativity, that, once augmented with the ``electromagnetic term" to become H$'_{1\ \mathrm{em}}$, leads to the correct energy levels for the electron in the hydrogen atom. Thus, suppose that $\Psi $ is an eigenstate, with energy $E$, of the special-relativistic Dirac Hamiltonian H$'_{1\ \mathrm{em}}$ for the electron in the hydrogen atom. Let $E'$ be the corresponding energy mean value got by using the DFW Hamiltonian H$'_{3\ \mathrm{em}}$ --- which is valid in the same inertial frame (the mass center frame) as is H$'_{1\ \mathrm{em}}$, but that uses Ryder's tetrad instead of the Cartesian tetrad. As shown by Eq. (\ref{Etilde-Ebreve-em}), we have $E'-E=A $, where $A$ is given by Eq. (\ref{bar A-explicit}): $A$ depends on the eigenstate $\Psi $ and is arbitrarily large.\\

Although Ryder's tetrad is not ``adapted" to the rotating frame F in the sense of Eq. (\ref{u_0=v_F}), it will turn out to be interesting to have the precise expression of the Hamiltonian and energy operator in that frame F [in the coordinates $(x^\mu )$, Eq. (\ref{rotating Cartesian})] with this tetrad. That precise expression was not given by Ryder \cite{Ryder2008}, who wrote: ``The Dirac equation (4) then, on rearrangement, is found to have a $\Mat{\sigma }.\Mat{\omega }$ ($=\omega \sigma ^3$ here) contribution to the Hamiltonian --- a spin-rotation coupling term exactly as predicted by Mashhoon." From the expression (\ref{Ryder1}) of that tetrad as function of the natural basis of the coordinates $(x^\mu )$, and from (\ref{Spin connec - Ryder1}), we get once again for the connection matrices [this time in the coordinates $(x^\mu )$]:
\be\label{Spin connec - Ryder1 - rotating Cartesian}
\Gamma_0= -\frac{i\omega}{2c}\, \Sigma ^3, \quad \Gamma_j=0,
\ee
and we get the $\gamma ^\mu $ matrices (\ref{flat-deformed}),
\be\label{gamma - Ryder1 - rotating Cartesian}
\gamma  ^0=\gamma ^{\sharp 0},\quad \gamma  ^1=\frac{\omega y}{c}\gamma ^{\sharp 0}+\gamma ^{\sharp 1},\quad \gamma ^2=-\frac{\omega x}{c}\gamma ^{\sharp 0}+\gamma ^{\sharp 2},\quad \gamma ^3=\gamma^{\sharp 3}
\ee
[thus once more the hermiticity condition (\ref{hermiticity-condition}) is verified], whence for the $\alpha ^\mu $ matrices in Eq. (\ref{alpha}):
\be\label{alpha - Ryder1 - rotating Cartesian}
\alpha^0=\gamma ^{\sharp 0},\quad \alpha ^1= \frac{\omega y}{c}{\bf 1}_4+\alpha  ^{\sharp 1}, \quad \alpha ^2= -\frac{\omega x}{c}{\bf 1}_4+\alpha  ^{\sharp 2}, \quad \alpha  ^3=\alpha^{\sharp 3}.
\ee
Therefore, the energy operator $\mathrm{E}_3 = \mathrm{H}_3$, Eq. (\ref{Hamilton-Dirac-normal}), is now:
\bea 
\mathrm{H}_3 & = & mc^2\alpha ^0 -i\hbar c \alpha ^{ j} \partial_j -i\hbar c \Gamma _0\nonumber\\
& = & mc^2\gamma ^{\sharp 0} -i\hbar c \left[\alpha ^{\sharp j}\partial _j +\frac{\omega }{c}(y\partial _x-x\partial _y) +\Gamma _0\right], \\
\mathrm{H}_3 & = & \mathrm{H}'_3 -\Mat{\omega .}{\bf L}.
\label{Hamilton-Ryder1-rotating}
\eea
Remembering Eq. (\ref{Spin connec - Ryder1 - rotating Cartesian})
, we see that with Ryder's (first) tetrad, the energy operator in the rotating frame F has indeed the spin-rotation coupling term $-\frac{\hbar \omega }{2}\Sigma ^3=-\Mat{\omega }.{\bf S }$ --- as has the energy operator with this tetrad but in the inertial frame F$'$, Eq. (\ref{Hamilton-Ryder1-inertial}). Also, these two energy operators differ from one another only by the {\it angular momentum} term --- just as we found also with the Cartesian tetrad, Eqs. (\ref{Hamilton-Dirac-SR}) and (\ref{Hamilton-restricted-gauge}).

\subsection{The general relation between the Hamiltonians in two frames in relative rotation}

It turns out to be a general fact that the Dirac Hamiltonian operators in two reference frames in relative rotation differ only by the angular momentum term, if they correspond to the same tetrad field. In a general Lorentzian spacetime $(\mathrm{V},\Mat{g})$, consider a general reference frame R$'$, defined by a chart $\chi' : X\mapsto (x'^\mu )=(ct',x',y',z')$, and define another reference frame R by a chart $\chi $ deduced from $\chi '$ by a transformation generalizing (\ref{rotating Cartesian}):
\be\label{rotating chart}
t=t',\ x=x'\cos \phi (t) + y' \sin \phi (t),\ y=-x' \sin \phi (t) + y' \cos \phi (t),\ z=z'.
\ee
So the spatial coordinate vector ${\bf r}\equiv (x,y,z)$, at least, is undergoing a rotation, at a variable rate $\omega \equiv \dot{\phi }\equiv d\phi /dt$, with respect to the space browsed by the coordinates $(x'^j)$. This corresponds to a  rotation in physical space if $\mathrm{V}$ is endowed with the Minkowski metric $\Mat{\gamma}$ (with possibly $\Mat{\gamma}=\Mat{g}$ as a particular case) and the chart $\chi' $ is Cartesian for $\Mat{\gamma}$. The Dirac Hamiltonian (\ref{Hamilton-Dirac-normal}) rewrites immediately, in the most general case, as
\be \label{Hamilton-DFW-space-covariant}
\mathrm{H} = i\frac{\partial}{\partial t}  + \frac{m}{g^{00}}\gamma ^0 -i\frac{\gamma ^0\gamma ^\mu  D_\mu}{g^{00}}   \qquad (\hbar =1=c).
\ee
On a general coordinate change, $\gamma ^\mu  D_\mu$ is invariant (for a given tetrad field, of course), due to the transformation behaviours of $\gamma ^\mu $, $\Gamma _\mu $, and $\partial _\mu $. On the coordinate change (\ref{rotating chart}), $\gamma ^0$ and $g^{00}$ are invariant. Therefore, we have
\be \label{Hamilton-change-rotation}
\mathrm{H}'- \mathrm{H} = i\left(\frac{\partial }{\partial t'}-\frac{\partial }{\partial t}\right) = i\frac{\partial x^j}{\partial t'} \frac{\partial }{\partial x^j}=i\,\dot{\phi }\left(y\frac{\partial }{\partial x}-x\frac{\partial }{\partial y}\right),
\ee
that is,
\be\label{delta H-relative rotation}
\mathrm{H} - \mathrm{H}'= -\Mat{\omega .}{\bf L}, \quad \Mat{\omega }\equiv (\omega,0,0), \quad {\bf L}\equiv {\bf r}\wedge (-i\hbar \nabla ),
\ee
as announced. 
\footnote{\
In particular, the Hamiltonian in the inertial frame F$'$ and with the adapted tetrad (\ref{u circ-01})-(\ref{u circ-23}) is: $\mathrm{H}'_2=\mathrm{H}_2+\Mat{\omega .}{\bf L}$, with $\mathrm{H}_2$ given by Eq. (\ref{Hamilton-Dirac-normal-2}). This is also the energy operator.
}

\section{Conclusion}

The predictions of the spin-rotation coupling term for a particle obeying the covariant Dirac equation (\ref{Dirac-normal}) have considered a tetrad field which is undergoing more or less the same rotation as the rotating reference frame itself \cite{HehlNi1990,CaiPapini1991,Ryder2008}. As suggested by Hehl \& Ni \cite{HehlNi1990} and by Ryder \cite{Ryder2008}, to make this precise one should use the notion of the Fermi-Walker transport or derivative. By using the Fermi-Walker derivative one may indeed define rigorously the rotation rate of the spatial triad associated with an orthonormal tetrad, for a general reference frame in a general spacetime  \cite{A47}. This rotation rate is the spatial tensor field $\Mat{\Xi }$ in Eq. (\ref{Xi=-Phi}). That definition needs that one considers an ``adapted" tetrad to the reference frame considered, i.e., one such that the time-like vector of the tetrad is the four-velocity of the reference frame, Eq. (\ref{u_0=v_F}). The rotation rate of the reference frame should be precisely defined also as a spatial tensor field $\Mat{\Omega  }$ and can indeed be, Eq. (\ref{Weyssenhoff modified}). \\

For a uniformly rotating frame in the Minkowski spacetime, we succeeded at defining an {\it adapted} tetrad field which verifies $\Mat{\Xi }=\Mat{\Omega  }$ almost exactly. With this tetrad field, the energy operator in the rotating frame does have the spin-rotation coupling term, Eq. (\ref{Spin-rotation-2}). We also wrote explicitly the energy operator with Ryder's rotating tetrad field \cite{Ryder2008}, which does involve this term, too --- although Ryder's tetrad is adapted to the inertial frame, not to the rotating frame.\\

However, the three tetrad fields investigated in the present work provide three different Hamiltonians in the inertial frame, as well as three different Hamiltonians in the rotating frame. (In each case, the Hamiltonian coincides with the energy operator.) We emphasized the grave physical inequivalence of the energy operators in the inertial frame and corresponding with either the Cartesian tetrad or Ryder's rotating tetrad. Moreover, those tetrads that provide the spin-rotation coupling term in the energy operator of the rotating frame, give it also in the energy operator of the {\it inertial frame}. In fact, we find quite generally that the Hamiltonian operators in two reference frames in relative rotation, but corresponding to the same tetrad field, differ {\it only} by the {\it angular momentum} term, Eq. (\ref{delta H-relative rotation}). Thus, if the Hamiltonian involves spin-rotation coupling in the rotating frame, and if one keeps the same tetrad, then the corresponding Hamiltonian in the inertial frame {\it must} also involve spin-rotation coupling, which is certainly unexpected physically. Therefore, if the spin-rotation coupling is to exist for a Dirac particle, it means that two different tetrad fields must be chosen for two different reference frames. Thus, for each given reference frame, a tetrad field adapted to that reference frame should be chosen. Then, to get the relevant rotation rate in the spin-rotation coupling term, one has to impose that the rotation rate of the triad is indeed that of the reference frame: $\Mat{\Xi }=\Mat{\Omega  }$. That is, if the spin-rotation coupling is to exist for a Dirac particle, Variant {\it a}) of \hyperref[Framework I]{Framework I} is the correct scheme to select the tetrad field. As we saw, this is difficult to implement already for the simple case of a uniform rotation in a Minkowski spacetime --- not to speak of a general situation.\\

One may consider that the choice of a tetrad field should be valid for any reference frame instead. \hyperref[Framework II]{Framework II} is the only currently available one that ensures this while providing unambiguous Dirac Hamiltonian and energy operators. It assumes that the metric can be put in the form (\ref{isotropic-diagonal}) in some chart: preferably a global one of course, in which case, by setting $\gamma _{\mu \nu }\equiv \eta _{\mu \nu }$ in that chart, one endows the spacetime with the Minkowski metric $\Mat{\gamma } $, related simply to the physical metric $\Mat{g}$. In the case of a Minkowski spacetime ($\Mat{g}=\Mat{\gamma } $), this framework leads to selecting any ``Cartesian tetrad". It predicts no spin-rotation coupling. Thus, experiments should decide.



\begin{thebibliography}{9}
\small

\bibitem{WernerStaudenmannColella1979}
S. A. Werner, J. L. Staudenmann, and R. Colella, ``Effect of Earth's rotation on the quantum mechanical phase of the neutron," {\it Phys. Rev. Lett.} {\bf 42}, 1103--1106 (1979).

\bibitem{A41}
M. Arminjon, ``Main effects of the Earth's rotation on the stationary states of ultra-cold neutrons," {\it Phys. Lett. A} {\bf 372}, 2196--2200 (2008). [See also \href{http://arxiv.org/abs/0708.3204v2}{arXiv:0708.3204v2 (quant-ph)}]

\bibitem{Kuroiwa-et-al1993}
J. Kuroiwa, M. Kasai, and T. Futamase, ``A treatment of general relativistic effects in quantum interference," {\it Phys. Lett. A} {\bf 182}, 330--334 (1993).

\bibitem{MorozovaAhmedov2009}
V. S. Morozova and B. J. Ahmedov, ``Quantum interference effects in slowly rotating NUT space-time," {\it Int. J. Mod. Phys. D} {\bf 18}, 107--118 (2009). \href{http://arxiv.org/abs/0804.2786v2}{[arXiv:0804.2786v2 (gr-qc)]}

\bibitem{Mashhoon1995}
B. Mashhoon, ``On the coupling of intrinsic spin with the rotation of the earth," {\it Phys. Lett. A} {\bf 198}, 9--13 (1995).

\bibitem{Mashhoon1988}
B. Mashhoon, ``Neutron interferometry in a rotating frame of reference," {\it Phys. Rev. Lett.} {\bf 61}, 2639--2642 (1988).

\bibitem{HehlNi1990}
F. W. Hehl and W. T. Ni, ``Inertial effects of a Dirac particle," {\it Phys.\ Rev.\  D} {\bf 42}, 2045--2048 (1990).

\bibitem{CaiPapini1991}
Y. Q. Cai and G. Papini, ``Neutrino helicity flip from gravity-spin coupling," {\it Phys. Rev. Lett.} {\bf 66}, 1259--1262 (1991).

\bibitem{BrillWheeler1957+Corr}
D. R. Brill and J. A. Wheeler, ``Interaction of neutrinos and gravitational fields," {\it Rev. Modern Phys.} {\bf 29}, 465--479 (1957). Erratum: {\it Rev. Modern Phys.} {\bf 33}, 623--624 (1961).

\bibitem{ChapmanLeiter1976}
T. C. Chapman and D. J. Leiter, ``On the generally covariant Dirac equation," {\it Am. J. Phys.} {\bf 44}, No. 9, 858--862 (1976).

\bibitem{Isham1978}
C. J. Isham, ``Spinor fields in four dimensional space-time," {\it Proc. Roy. Soc. London A} {\bf 364}, 591--599 (1978).

\bibitem{Ryder2008}
L. Ryder, ``Spin-rotation coupling and Fermi-Walker transport," {\it Gen. Relativ. Gravit.} {\bf 40}, 1111--1115 (2008).

\bibitem{A43}
M. Arminjon and F. Reifler, ``A non-uniqueness problem of the Dirac theory in a curved spacetime," {\it Ann. Phys. (Berlin)} {\bf 523}, 531--551 (2011). \href{http://arxiv.org/abs/0905.3686}{[arXiv:0905.3686 (gr-qc)]}

\bibitem{A45}
M. Arminjon and F. Reifler, ``Four-vector vs. four-scalar representation of the Dirac wave function," {\it Int. J. Geom. Meth. Mod. Phys.} {\bf 9}, No. 4, 1250026 (2012). \href{http://arxiv.org/abs/1012.2327v2}{[arXiv:1012.2327v2 (gr-qc)]}

\bibitem{Leclerc2006}
M. Leclerc, ``Hermitian Dirac Hamiltonian in the time-dependent gravitational field," {\it Class. Quant. Grav.} {\bf 23}, 4013--4020 (2006). \href{http://arxiv.org/abs/gr-qc/0511060v3}{[arXiv:gr-qc/0511060v3]}

\bibitem{A48}
M. Arminjon, ``A simpler solution of the non-uniqueness problem of the Dirac theory," {\it Int. J. Geom. Meth. Mod. Phys.} {\bf 10}, No. 7, 1350027 (2013) [24 pages]. \href{http://arxiv.org/abs/1205.3386v4}{[arXiv:1205.3386v4 (math-ph)]}

\bibitem{GorbatenkoNeznamov2013}
M. V. Gorbatenko and V. P. Neznamov, ``Absence of the non-uniqueness problem of the Dirac theory in a curved spacetime. Spin-rotation coupling is not physically relevant," \href{http://arxiv.org/abs/1301.7599v2}{arXiv:1301.7599v2 (gr-qc)}.

\bibitem{A50}
M. Arminjon, ``On the non-uniqueness problem of the covariant Dirac theory and the spin-rotation coupling," {\it Int. J. Theor. Phys.} {\bf 52} (2013), DOI 10.1007/s10773-013-1717-x. [\href{http://arxiv.org/abs/1302.5584v2}{arXiv:1302.5584v2 (gr-qc)}]

\bibitem{Parker1980}
L. Parker, ``One-electron atom as a probe of spacetime curvature," {\it Phys. Rev. D} {\bf 22}, 1922--1934 (1980).

\bibitem{HuangParker2009}
X. Huang and L. Parker, ``Hermiticity of the Dirac Hamiltonian in curved spacetime," {\it Phys. Rev. D} {\bf 79}, 024020 (2009). \href{http://arxiv.org/abs/0811.2296}{[arXiv:0811.2296 (gr-qc)]}

\bibitem{A47}
M. Arminjon, ``A solution of the non-uniqueness problem of the Dirac Hamiltonian and energy operators," {\it Ann. Phys. (Berlin)} {\bf 523}, 1008--1028 (2011). [Pre-peer-review version: \href{http://arxiv.org/abs/1107.4556v2}{arXiv:1107.4556v2 (gr-qc)}].

\bibitem{A40}
M. Arminjon and F. Reifler, ``Dirac equation: Representation independence and tensor transformation," {\it Braz. J. Phys.} {\bf 38}, 248--258 (2008). \href{http://arxiv.org/abs/0707.1829}{[arXiv:0707.1829 (quant-ph)]}


\bibitem{Cattaneo1958}
C. Cattaneo, ``General relativity: relative standard mass, momentum, energy and gravitational field in a general system of reference," {\it il Nuovo Cimento} {\bf 10}, 318--337  (1958).

\bibitem{Weyssenhoff1937}
J. von Weyssenhof, ``Metrisches Feld und Gravitationsfeld," {\it Bull. Acad. Polon. Sci., Sect. A} {\bf 252} (1937). (Quoted by Cattaneo \cite{Cattaneo1958}.)

\bibitem{A42}
M. Arminjon and F. Reifler, ``Basic quantum mechanics for three Dirac equations in a curved spacetime," {\it Braz. J. Phys.} {\bf 40}, 242--255 (2010). \href{http://arxiv.org/abs/0807.0570}{[arXiv:0807.0570 (gr-qc)]}.

\bibitem{ChernChenLam1999}
S. S. Chern, W. H. Chen, and K. S. Lam, {\it Lectures on Differential Geometry} (Singapore: World Scientific 1999), pp.  113--121.

\bibitem{Pauli1936}
W. Pauli, ``Contributions math\'ematiques \`a la th\'eorie des matrices de Dirac," {\it Ann. Inst. Henri Poincar\'e} {\bf 6}, 109--136 (1936).

\bibitem{MashhoonMuench2002}
B. Mashhoon and U. Muench, ``Length measurement in accelerated systems," {\it Ann. Phys. (Berlin)} {\bf 11}, 532--547 (2002). \href{http://arxiv.org/abs/gr-qc/0206082v1}{[arXiv:gr-qc/0206082v1]}

\bibitem{MalufFariaUlhoa2007}
J. W. Maluf, F. F. Faria, and S. C. Ulhoa, ``On reference frames in spacetime and gravitational energy in freely falling frames," {\it Class. Quant. Grav.} {\bf 24}, 2743--2754 (2007). \href{http://arxiv.org/abs/0704.0986v1}{[arXiv:0704.0986v1 (gr-qc)]}

\bibitem{A35}
M. Arminjon, ``Space isotropy and weak equivalence principle in a scalar theory of gravity," {\it Braz. J. Phys.} {\bf 36}, 177--189 (2006). \href{http://arxiv.org/abs/gr-qc/0412085} {[arXiv:gr-qc/0412085]}

\bibitem{A44}
M. Arminjon and F. Reifler, ``General reference frames and their associated space manifolds," {\it Int. J. Geom. Methods Mod. Phys.} {\bf 8}, No. 1, 155--165 (2011). [\href{http://arxiv.org/abs/1003.3521}{arXiv:1003.3521v2 (gr-qc)}]


\bibitem{JantzenCariniBini1992}
R. T. Jantzen, P. Carini, and D. Bini, ``The many faces of gravitoelectromagnetism," {\it Ann. Phys. (New York)} {\bf 215}, 1--50 (1992). \href{http://arxiv.org/abs/gr-qc/0106043}{[arXiv:gr-qc/0106043]}






\end{thebibliography}
\end{document}